\documentclass[10pt, conference, final, letterpaper]{IEEEtran}

\usepackage{float}
\usepackage{verbatim}
\usepackage{algorithm}
\usepackage[noend]{algpseudocode}
\usepackage{textcmds}
\usepackage{url}
\usepackage{amsmath}
\usepackage{wasysym}
\usepackage{graphicx}
\usepackage{fixltx2e}
\usepackage{subfig}
\usepackage{amsthm}

\theoremstyle{definition}
\newtheorem{defn}{Definition}[section]

\makeatletter
\renewcommand{\ALG@beginalgorithmic}{\scriptsize}
\makeatother

\IEEEoverridecommandlockouts
\author{
\IEEEauthorblockN{Yang CHI
{} and Dharma P. Agrawal
}
\IEEEauthorblockA{
Center of Distributed and Mobile Computing, School of Computing Sciences and Informatics\\
University of Cincinnati, Cincinnati, OH 45221 - 0008}
Email: chiyg@mail.uc.edu, dpa@cs.uc.edu
\thanks{Yang Chi was with University of Cincinnati when this work was done. He is currently with Cisco Systems, Inc., and is reacheable via yangchi@cisco.com.}
}

\title{TCP-Forward: Fast and Reliable TCP Variant for Wireless Networks}

\begin{document}
\maketitle

\begin{abstract}
The congestion control algorithms in TCP may incur inferior performance in a lossy network context like wireless networks. Previous works have shown that random linear network coding can improve the throughput of TCP in such networks, although it introduces extra decoding delay at the destination. In this paper we try to alleviate the decoding delay by replacing random linear network coding with LT Codes. Due to the inherent difference between linear network coding and Fountain Codes, such replacement is not as simple as it sounds. We conquer some practical problems and come up with TCP-Forward, a new TCP variant which offers many properties that TCP as a streaming transport protocol should offer. Our performance evaluation shows TCP-Forward provides better performance than previous works.
\end{abstract}

\section{Introduction}
\label{sec:intro}

The most common wireless networks technology, IEEE 802.11 specifications, were first introduced in the late 1990s \cite{IEEE80211Legacy} \cite{IEEE80211Legacy99}, when most protocols in the TCP/IP stack had been existing and evolving for more than a decade. The widespread belief is a hierarchical model of network protocol stack can help hiding details from layer to layer. It is also well accepted that the design of one layer generally should not have to consider the implementation details of underlying layers. This, however, is not the case for TCP for have a reasonable performance in a wireless context.

The problem with using TCP in wireless networks arises from both protocols' inherent nature: wireless medium has a higher lossy percentage and the signal is way more error-prone during transmission, compared to wired networks. On the other hand, TCP considers packet loss or packet drop as an indicator of congestion. In fact, TCP assumes congestion is the only reason for packet loss. This has worked very well in the wired networks, since packet loss or transmission error caused by the Ethernet or other wired MAC layer technologies themselves is very rare. However, the packet loss rate during wireless transmission is much higher, and is subject to surrounding physical environments, devices and traffics. When such packet loss happens, TCP will mistakenly see it as a signal of congestion, and will unnecessarily slow down its sending rate. Such wrongfully and immature reaction to packet loss is the key factor in TCP's performance drop in wireless networks.

Many solutions have been proposed to this problem. We divide them into two categories. The first one is new TCP congestion control mechanisms \cite{CongestionControl}. The philosophy behind this methodology is very simple. The problem arises from the conflict between the TCP and the wireless medium. Since we cannot change the wireless medium, the TCP congestion control mechanism simply becomes the most natural place to implement changes. Such solutions are more orthodoxy in the sense that they try to adopt most of the existing components and algorithms in TCP, stick to the end-to-end principle \cite{end2end}, and solve the problem from the sender. Plus, they try to achieve the services that TCP is supposed to provide (a reliability, connection-oriented and stateful service). They try to solve the problem within the TCP, and without introducing new protocols into the TCP/IP stack, or changing existing layers or protocols. 

Take TCP-Vegas \cite{TCP-Vegas} as an example. TCP-Vegas keeps record of the sending time of every packet it sends out and the time it receives the acknowledgement (ACK). This accurately measured RTT is the core of this protocol. TCP-Vegas keeps congestion avoidance and slow start algorithms from TCP-Tahoe and TCP-Reno. However, the condition on which the protocol transit from one state (e.g. congestion avoidance, slow start, etc.) to another is different. Retransmission, congestion avoidance and slow start all depend on the RTT value and its fluctuation. Duplicate ACKs (DUPACK) are not enough to make the protocol transit from one state to another one in its Finite State Machine (FSM). Other TCP variants that are proposed to solve the poor performance that TCP has in wireless networks use similar philosophy to avoid to overreact to DUPACKs. One problem with such methodology is that when such RTT-based TCP variants co-exists with DUPACK-based TCP variants (TCP-Tahoe, TCP-Reno, TCP-NewReno \cite{NewReno-RFC} \cite{NewReno-Paper}, TCP-BIC \cite{TCP-BIC}, TCP-Cubic \cite{TCP-Cubic}), RTT-based TCP variants often reduce its sending rate too early, as RTT fluctuation is a more sensory trigger compared to the DUPACKs. After they reduce the sending rate, the bottleneck node would have more bandwidth, and the DUPACK-based TCPs would simply take those bandwidth very quickly, becuase for them, there is no congestion and they have been increasing their sending rate during the whole time. Such unfairness problem are discussed in detailed by La et al. \cite{VegasIssue}.

Another commonly used TCP hacking to improve its performance in the wireless networks is to use the selective acknowledgements (SACK) option \cite{SACK-RFC}. The cumulative acknowledgements used by default in TCP does not provide sufficient information to the sender so it can react quickly when packet loss or congestion happens. With SACK, the sender will also be informed noncontiguous data that are also received by the receiver. Unfortunately, there are previous results showing that SACK mechanism is not able to improve TCP performance in lossy or error-prone networks effectively, compared to cumulative acknowledgements \cite{TCPWireless}.

The other category are the solutions that try to solve the problem outside of TCP, or even change the architecture of the network. For example, there are proposals that suggest replacing CSMA/CA by more reliable link-layer technologies like CDMA or TDMA altogether. Apparently, such thinking simply ignores the issue at all. Unfortunately, they would hardly work in reality. Moreover, contention-based MAC layer protocols like CSMA/CA essentially offers different service than CDMA and TDMA. Using CDMA and TDMA to replace CSMA/CA simply means the same service will not be fulfilled. 

Other mechanisms rely on the base stations to interfere with the transmission. One way is to introduce Explicit Loss Notification (ELN) at the base station. The base station keeps track of lost packets, and when DUPACK comes from the receiver, base station will modify the TCP header and make the sender informed that the packet loss is due to the wireless transmission. A similar mechanism is adopted in Snoop \cite{Snoop} protocol. Different from ELN, Snoop is used when the link between receiver and the base station is wireless. A Snoop agent has to be installed on the base station. It does not just keep track of, but also caches by-passing packets towards the wireless receiver. It detects packet loss by observing the DUPACKs. When it happens, instead of forwarding the DUPACKs to the sender, it retransmits the missing packet from its cache, and drops the DUPACKs.

The Split Connection protocols \cite{I-TCP} \cite{Split-MTCP} go even further by splitting the TCP connection into two at the bast station: one between the sender and the base station, and another one between the receiver and the base station. For the wired connection, regular TCP is used. For the wireless TCP connection, however, some special tuning of TCP or even some other transport layer protocol would be used.

Similar to the SACK approach, previous experiments have already indicated that these base station based mechanisms actually cannot improve the performance effectively \cite{TCPWireless}. Even worse, they make the base station very complex since they need to maintain information for every TCP connection. They violate the end-to-end principle, and the senders no longer have a correct information of the receiver. For example, in Split Connection protocols, the base station would acknowledge the receiving of a packet even before the packet reaches the real receiver. 
Since end-to-end connection is broken down by such protocols, end-to-end encryption and decryption schemes like TLS \cite{TLS-RFC} also become infeasible. 
The biggest problem of these mechanism for us, however, is that they assume a single-hop wireless network. These mechanisms would not even work in a multi-hop scenario.

To sum it up, most previous solutions to this problem are not very promising, both from performance perspective and practicability and usability perspective. A better solution is certainly called for. 

\section{TCP/NC: Applying Network Coding in TCP}
\label{sec:TCPNC}

Now, let the network coding save us from the TCP problem as well. To see how this makes sense, take a look at the following example in Figure \ref{fig:reliability}: 

\begin{figure}[hbt]
	\begin{center}
		\includegraphics[width=2.5in]{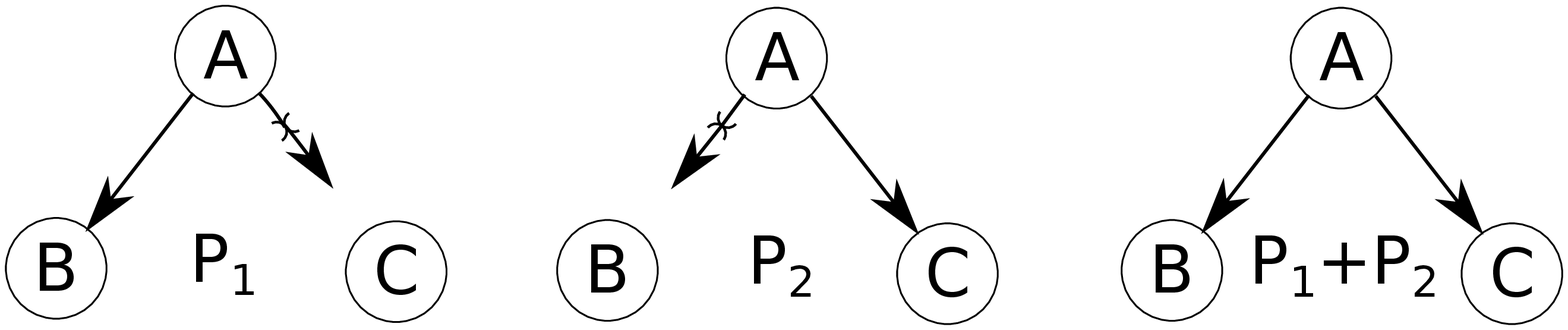}
	\end{center}
	\caption{How network coding helps with improving reliability in wireless networks. \cite{TCPNC}}
	\label{fig:reliability}
\end{figure}

In this example, node A needs to send both packet $P_{1}$ and packet $P_{2}$ to both node B and node C. Unfortunately, due to packet loss or transmission error, node C does not receive $P_{1}$ during the first broadcast and node B does not receive $P_{2}$ during the second broadcast. Without network coding, A has to broadcast both packets again, in two separate transmissions. Thanks to network coding, node A actually only has to broadcast a linear combination of these packets instead, and both B and C will be able to recover the packets from the combination.

\begin{figure}[hbt]
	\begin{center}
		\includegraphics[width=1.5in]{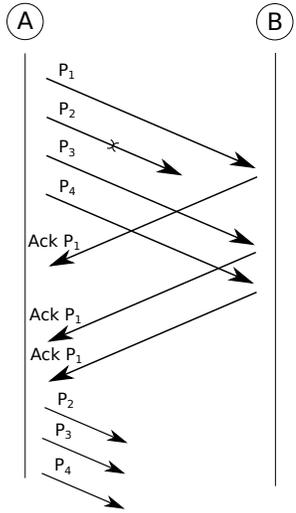}
	\end{center}
	\caption{End-to-end reliable transmission with ARQ.}
	\label{fig:ARQ}
\end{figure}

\begin{figure}[hbt]
	\begin{center}
		\includegraphics[width=2in]{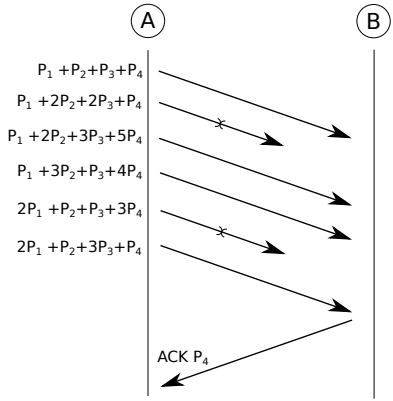}
	\end{center}
	\caption{End-to-end reliable transmission with linear coding.}
	\label{fig:FEC}
\end{figure}

Network coding's capacity of improving reliability does not limit to the broadcast or multicast situations. In unicast, we can achieve the same thing. Suppose node A needs to send node B a series of packets. Instead of sending the plain packets, A can send B a series of independent linear combinations of these packets. If the total number of packets to be sent is $N$, then as long as $N$ independent linear combinations are received by the receiver, the receiver can decode and recover the original packets. If the packet loss ratio is $0 < P_{loss} < 1$, then the total number for the sender to send out is $N \over 1 - P_{loss}$ for the receiver to receive $N$ packets. What's more, with this scheme, no Automatic Repeat Request (ARQ, i.e., using explicit acknowledgements to notify the sender that the last packet is received and it is OK to send the next packet) is required for each single packet. To see why is this particularly helpful, take a look at the example in Figure \ref{fig:ARQ} and Figure \ref{fig:FEC} where node A need to send 4 packets to node B. Suppose packet $P_{2}$ is lost during transmission. With ARQ, node A will only know this until it receives duplication acknowledgements for packet $P_{1}$. Then node A not just has to retransmit packet $P_{2}$, but also packet $P_{3}$ and $P_{4}$. Whereas with network coding, the sender just has to keep generating and sending out independent linear combinations. When the receiver receives $4$ such combinations, it will decode them, and acknowledge all of them. Note than Figure \ref{fig:ARQ} is exactly how TCP works, and Figure \ref{fig:FEC} is exactly why network coding can help us out of this troublesome.

Built upon the same idea, TCP/NC \cite{TCPNC} becomes the first work that demonstrate network coding's usability of improving TCP performance in the wireless contexts. More importantly, this work proves the effectiveness of Forward Error Coding (FEC, Linear Network Coding, e.g. Figure \ref{fig:FEC}, being one instance) over ARQ in a end-to-end settings.

\section{TCP-Forward: Fast and Reliable TCP Variant for Wireless Networks}
\label{chapter:TCPLT}

The biggest problem of TCP/NC is the decoding delay. In Figure \ref{fig:FEC}, decoding operations is only possible after node B receives 4 linear combinations. Only at that point, TCP/NC can start to decode and forward decoded packets to the upper layer applications. Another drawback with TCP/NC is the decoding complexity. Random Linear Network Coding relies on Gaussian Elimination, which is a complex algorithm, to decode an encoded packet. This, again, makes a longer decoding delay, and requires higher computational capacity and energy consumption. In this chapter, we address these two problems by introducing TCP-Forward, a new transport layer protocol that replace Random Linear Network Coding with Fountain Codes.

Fountain Codes was first proposed in the late 1990s \cite{fountain}. M. Luby's proposal of LT Codes \cite{ltcodes} became the first practical realization of Fountain Codes and the foundation of most following works in Fountain Codes. The general idea behind Fountain Codes is very similar to Random Linear Network Coding: instead of using a feedback channel to notify the source if the sent data successfully arrives at the destination, redundancy is introduced to make sure the destination node can get the original data even if the transmission channel is lossy. What makes Fountain Codes different from other Erasure Codes are the two properties it possesses. First, Fountain Codes are not generated at a fixed rate. This is why Fountain Codes is also known as Rateless Erasure Codes. Secondly, given a finite sequence of input data, Fountain Codes is capable of generating an infinite sequence of encoded symbols. These two properties make it very feasible to utilize Fountain Codes in practical networking protocols. Different kinds of Fountain Codes have been used in many wireless communication standards including IEEE 802.11n, CDMA2000, EV-DO, 3GPP and even 10GBase-T Ethernet. A key observation is that such techniques have been mainly used in single-hop communications. What we want to achieve, is to use Fountain Codes in end-to-end communication.

\section{TCP-Forward: an introduction}
\label{section:TCP-Forward}

We take TCP-Vegas as the foundation of our protocol. The TCP-Vegas' congestion control algorithm is kept. A few other components are introduced in the TCP-Forward. Besides the regular TCP congestion window, the sender maintains another coding window for the Fountain Codes encoding operations. In this dissertation we will use LT Codes as our encoding/decoding algorithm, although other Fountain Codes like Raptor Codes can just work as well. The reason that we pick LT Codes is that it is not just one particular Fountain Codes realization, but also the core algorithm of the following Fountain Codes algorithms including Raptor Codes and Online Codes. As a result, we can easily ``plug-n-play'' other Fountain Codes algorithms into our protocol once we finish design it with LT Codes. The receiver takes care of packet decoding. The intermediate nodes do not have to be aware of the existence of the coding algorithms at the transport layer and do not have to take part into encoding or decoding. The path between them include one or more hops of links, each of which has a positive packet loss rate. Redundancy is introduced at the sender to provide reliability, but explicit acknowledgements are still sent by the receiver. However, different from regular TCP congestion control algorithms, this acknowledgement is only used to move the coding window, which in turn slides the TCP congestion window. Duplicate acknowledgements will not incur TCP to reduce its transmission rate, nor will it change any parameters used in the congestion avoidance algorithms in TCP. In the rest of this chapter, we will provide more details of this new TCP variant. 

\subsection{TCP-Forward Sender}
\label{section:TCP-Forward:sender}

\begin{figure}[htb]
	\begin{center}
		\includegraphics[width=2.5in]{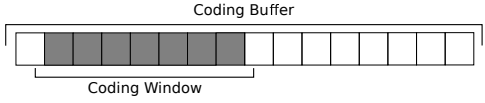}
	\end{center}
	\caption{The coding buffer and the coding window.}
	\label{fig:coding_window}
\end{figure}

When there are data that are ready to be sent in the TCP congestion window, they will be pushed into a coding buffer first \footnote{Note that the coding buffer and coding window are different.}. A various length window, a.k.a, the coding window as shown in Figure \ref{fig:coding_window}, is maintained to identify which segments can be encoded together in the current encoding round. 
The group of segments in the coding window is also sometimes referred to as a \textit{generation} of segments. During the encoding process, only the segments in the coding window will be used. Each round, $d$ segments from the coding window will be encoded together with bitwise XOR operation, where $1 \le d \le N$. The value of $d$ is thus also referred to as the \textit{degree} of the encoded symbol. When generating the value of $d$ in each encoding round, one needs to guarantee two things. First, every segment in the coding window has to be included in some encoded symbol at least once. Second, there is at least one encoded symbol with degree $d = 1$, so that the receiver can decode the packets. M. Luby first proposed a degree distribution that can guarantee such properties if the transmission channel is reliable \cite{ltcodes}. For readers' convenience, we briefly repeat it in the following paragraph.

This distribution, known as Ideal Soliton distribution, is defined in Definition \ref{def:ideal_soliton}. In practice, however, encoded symbols may get lost during transmission, which will make the decoding fail. Thus, Robust Soliton distribution, as in Definition \ref{def:robust_soliton}, is introduced to solve this problem. Simply speaking, Robust Soliton distribution is derived and modified from Ideal Soliton distribution to make sure that more encoded symbols will have a very small degrees (close to 1), and encoded symbols with larger than 1 degree will be few. During the encoding process, in case the segments in the coding window have different sizes, a series of $0$ is padded to the end of them to make them the same length.

\begin{defn}[Ideal Soliton distribution \cite{ltcodes}]\label{def:ideal_soliton}
	\begin{align}
	\rho(i) = \left\{
		\begin{array}{l l}
			1/k & \quad \text{for $i = 1$} \\
			1/{i(i-1)} & \quad \text{for $i = 2, \dots, k$}
		\end{array} \right.
	\end{align}
\end{defn}

\begin{defn}[Robust Soliton distribution \cite{ltcodes}]\label{def:robust_soliton}
	Let $R = c \cdot ln(k/\delta)\sqrt{k}$ for some constant $c > 0$. Define \\
	\begin{align}
		\tau(i) = \left\{
			\begin{array}{l l}
				R/ik & \quad \text{for $i = 1,\dots,k/R-1$} \\
				Rln(R/\delta)/k & \quad \text{for $i = k/R$} \\
				0 & \quad \text{for $i = k/R+1, \dots ,k$}
		\end{array} \right.
	\end{align}
	And let $\beta = \sum_{i = 1}^{k} \rho(i) + \tau(i)$, where $\rho(\cdot)$ is the Ideal Soliton distribution defined in Definition \ref{def:ideal_soliton}. Define Robust Soliton distribution $\mu(\cdot)$ as:
	\begin{align}
		\mu(i) = (\rho(i) + \tau(i))/\beta
	\end{align}
\end{defn}

For the receiver to be able to decode the packets, at least $N$ encoded symbols have to be successfully received at the destination. If the transmission channel is lossless, then only $N$ rounds of the encoding process described above need to perform. However, since we are dealing with lossy wireless links, a key factor would be introducing redundancy in the sent data. An ideal approach of calculating the redundancy factor would be to assume that the packet loss rate along each link is known \textit{a priori}. This assumption is not extremely audacious given that some routing protocols in the multi-hop wireless networks have to establish a path before forwarding data. Hence, during the path establishment and advertisement, the packet loss rates along the path can be piggy-backed to the sender, so that sender will be able to calculate an ideal redundancy factor. Nevertheless, as an end-to-end protocol, TCP-Forward chooses another approach to estimate the redundancy factor, which will be explained in later sections after acknowledgement is covered. For now, the readers can just remember that given $N$ segments, $N \times R$ encoded symbols are generated, where $R > 1$.

\begin{figure}[hbt]
	\begin{center}
		\includegraphics[width=2.5in]{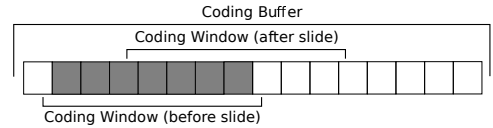}
	\end{center}
	\caption{The coding buffer and the coding window: sliding}
	\label{fig:coding_window_slide}
\end{figure}

The generated encoded symbols, although ready to be sent out, will be held temporarily in another buffer before they can be pushed down to the IP layer. The reason is to make sure the network is not congested. We adapt TCP-Vegas' congestion control algorithm for the same purpose. To make sure TCP-Forward would have the same sending rate behavior as TCP-Vegas, the sending buffer will only release packets to the network at a rate that is proportional to the rate that the TCP congestion window pushes segments to the coding buffer. To be more accurate, for every $K$ packets that TCP congestion window pushed to the coding buffer, the sending buffer will release $K \times R$ encoded symbols to the network, where $R$ is the same redundancy factor explained previously. One can understand the rate control in TCP-Forward is \textit{credit} based: every time a segment is pushed from TCP congestion window to the coding buffer, $R$ credits are earned. Sending buffer only pushes data to IP layer when the credit is positive. Every time an encoded symbol is pushed from sending buffer to IP layer, $1$ credit is cost.

After generating $N \times R$ encoded packets, the coding window will slide $\lfloor N/2 \rfloor$ segments as shown in Figure \ref{fig:coding_window_slide}. From there, another $N$ segments will be encoded into $N \times R$ encoded symbols and will be pushed to the sending buffer. The sliding windows are overlapped to further provide reliability.

\subsection{TCP-Forward Receiver and Data Acknowledgements}
\label{section:TCP-Forward:receiver}

The decoding algorithm is much simpler than the encoding algorithm, and is also way simpler and faster than the Gaussian Elimination which is used to decode Random Linear Network Coding. It starts by looking for an encoded symbol with degree $d = 1$. Once such a packet is identified, it will be bitwise XOR-ed with all the other encoded symbols in which it is included. This process will be repeated until either every symbol is decoded, or until there is no more encoded symbol with degree $d = 1$. The latter is a decoding failure, and will be handled by data acknowledgement (or more accurately, the lack of acknowledgement) and retransmission. 

What really gets tricky is not incorporating a decoding algorithm in TCP, but how to change the way TCP acknowledges a packet after the incorporation. The order of the original segments get decoded has a great chance to be different from the their original order, i.e., the segments with larger sequence number may get decoded before those with smaller sequence numbers. When this happens, the data will be kept temporarily in a receiving buffer before they can be forwarded to the application that is expecting the data to come in the correct order. 
An acknowledgement will be generated and sent back to the sender immediately after a segment gets decoded. Note that the acknowledgements in TCP-Forward are neither in-order not cumulative. When a segments gets decoded, an acknowledgement will be sent to the sender. This acknowledgement will make the original segment removed from the coding buffer at the sender. Then again, the coding function at the sender will generate an acknowledgement to the TCP congestion control algorithm. However, this newly generated acknowledgement will contain the sequence number of the first segment in the congestion window. Consequently, the congestion window will slide forward and its size will increase according to the congestion control algorithm. Without this sequence number translation process, the regular TCP will see an out-of-order acknowledgement and could simply drop it.

\begin{figure}[htb]
	\begin{center}
		\includegraphics[width=2.5in]{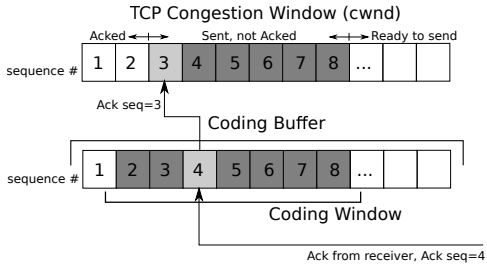}
	\end{center}
	\caption{The two-step acknowledgement and sequence number translation at the sender.}
	\label{fig:acks}
\end{figure}

Figure \ref{fig:acks} illustrates the above process. The receiver sends back an acknowledgement indicating that segment with sequence number $seq = 4$ is decoded. Upon receiving this, the sender will first remove this segment from its coding buffer. After that, it will modify the sequence number in the acknowledgement to be the smallest sequence number of a segment in the congestion window, in this case, $seq = 3$. Then, it forwards this modified acknowledgement to the regular TCP acknowledgment handler function, which will move the congestion window forward and enlarge the congestion window size so that more packets can be push to the coding buffer. The reason for this complex acknowledgment design is in two folds. First, the congestion control algorithm adapted from TCP-Vegas relies on the RTT to determine if the network is congested. If the receiver waits until segment with $seq = 2$ gets decoded to send back an acknowledgement, it may take such a long time that the sender will mistakenly decrease its congestion window. The second reason is, when we forward the acknowledgement from the coding function to regular TCP acknowledgement handler function at the sender, if we do not translate the sequence number in the acknowledgement packet, the every segment with a smaller sequence number will be mistakenly removed from the congestion window, due to the way acknowledgements are handled in conventional TCP.

\subsection{Retransmission, Redundancy Ratio and Other Details}

In TCP-Vegas, the congestion avoidance and slow-start only depends on the calculated RTT. The same mechanism can be borrowed by TCP-Forward. However, when it comes to retransmission mechanism, TCP-Forward has to handle it in a different way. In TCP-Vegas, DUPACK, though not a trigger of congestion windows size drop, is still one of the triggers of retransmission. However, in TCP-Forward, DUPACKs do not make sense, since the acknowledgements are not supposed to be in order anyway. And receiver will not send DUPACK when packet drop happens, simple because when it happens, the receiver will not be able to proceed the decoding process. For these reasons, DUPACKs are not viable in TCP-Forward as a packet drop indicator. Instead, we use the time that a segments spends in the coding window as the trigger. Similar to the way TCP-Vegas maintains an accurate estimate of RTT, for each original segment we keep track of the time it stays in the coding window. Let $ElapsedTime_{p}$ be the time segment packet $p$ spends in the coding window. Then define the following value $Diff_{p}$ for each packet $p$.:
\begin{align}
	\textit{Diff}_{p} = \frac{\text{coding window size}}{min_{i \in \text{all packets}}\{ElapsedTime_{i}\}} - \frac{\text{size of sigment p}}{ElapsedTime_{p}}.
\end{align}

When the value of $\textit{Diff}$ exceeds a predefined threshold \textit{Timeout} value, the TCP-Forward will start to generate encoded packets with all segments in the coding buffer but before the coding window. The encoded symbols will be sent to IP layer directly, instead of the sending buffer.

TCP-Forward also maintains two counters. One is the total number of TCP segments that have ever pushed into the coding window, denoted as $N_{segments}$. The other one is how many times the $\textit{Diff}$ value exceeds the $Timeout$ value, denoted as $N_{timeout}$. Then the redundancy ratio in TCP-Forward is defined as their ratio: $R = \frac{N_{segments}}{N_{segments} - N_{timeout}}$.

\section{Implementation and Experiment Results}
\label{sec:TCP-Forward:experiment}

We have implemented TCP-Forward on Linux system in the user space. We choose to implement it over UDP for the simplicity, although an ideal implementation would be using RAW socket to implement it over IP layer, or directly in the kernel space. Nevertheless, the current implementation is enough for the benchmark purpose.

We implement a simple sliding window algorithm over UDP to emulate TCP. Both Random Linear Network Coding and LT Codes are implemented, so we can have a rough comparison of TCP/NC and TCP-Forward. Our results are still preliminary at this point with some details of both protocols missing. That being said, the current implementations are enough to demonstrate the decoding performance of both protocols. We conducted both experiments and simulations in this work. The experimental network is quite small: only 3 nodes are included. A simulated network with lossy channels is implemented for the simulation. To demonstrate the advantage of TCP-Forward over TCP/NC in resource limited devices, we run both experiments and simulations in different platforms, including 32-bit Linux on a 32-bit Intel CPU, 64-bit Linux on a 64-bit Intel CPU, and a Raspberry Pi \cite{rpi} single board computer running 32-bit ARM Linux with a 700MHz ARM CPU. The 32-bit Intel CPU is quite out-dated in today's hardware standard, but still much faster than the Raspberry Pi. Table \ref{table:TCP:mems} shows the memory usage from experiments of both TCP-Forward and TCP/NC in these systems. Apparently memory usage should not be an issue for both protocols based on these results.

\begin{table*}[bht]
	\centering
\caption{System Memory Usage (RSS/VM Size in bytes) of TCP-Forward and TCP/NC}
\begin{tabular}{|c|c|c|c|} \hline
	&32-bit Intel+Linux &64-bit Intel+Linux &Raspberry Pi\\ \hline
	TCP-Forward&1283/3716&1476/16876&1360/3360\\ \hline
	TCP/NC&1508/4020&1541/17028&1512/3680 \\ \hline
\end{tabular}
\label{table:TCP:mems}
\end{table*}

\begin{figure}[htb]
	\begin{center}
		\includegraphics[width=2.5in]{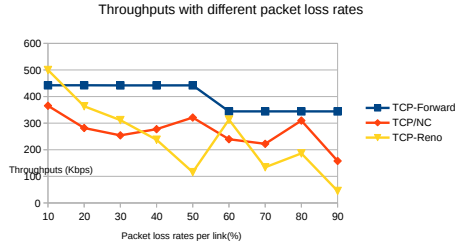}
	\end{center}
	\caption{TCP-Forward suffers least throughput drop with higher packet loss rate.}
	\label{fig:tcpforward:throughputs}
\end{figure}

Figure \ref{fig:tcpforward:throughputs} shows the throughputs of TCP-Forward, TCP/NC and TCP-Reno with different packet loss rate along the path. When the packet loss rate is low, TCP-Reno actually has better throughput than coded TCP (TCP/NC and TCP-Forward). As the packet loss rate increases, however, both TCP-Forward and TCP/NC start to show their strength. Between these two schemes, TCP-Forward has a very clear advantage, as it provides better throughputs with all different packet loss rates, and its throughput does not drop much with the reliability of the path gets extremely bad (packet loss rate at $90\%$).

What intrigues us more is the latency performance, since that is the major motivation behind this work. Therefore, we did additional simulations to investigate the latency from TCP-Forward and compared it to TCP/NC.

\begin{figure}[htb]
	\begin{center}
		\includegraphics[width=2.5in]{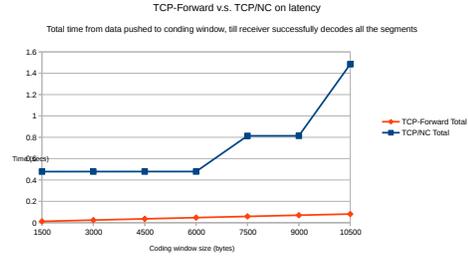}
	\end{center}
	\caption{The end-to-end latency from both TCP-Forward and TCP/NC.}
	\label{fig:tcpforward:tcpnc}
\end{figure}

Figure \ref{fig:tcpforward:tcpnc} illustrate end-to-end latency of both the TCP-Forward and TCP/NC schemes for a small simulated 3-node tandem network. This end-to-end latency include encoding delay, transfer delay (queueing, transmission and propagation delays combined), and decoding delay. TCP-Forward performs significantly better than TCP/NC based on our results, especially when the coding window gets larger. Fountain Codes, by its nature, should enjoy such an easy win over the Linear Coding. Actually, in the original TCP/NC paper \cite{TCPNC}, the simulated coding window size was never larger than $5$ TCP segments. In their simulation, TCP/NC performs best when the coding window size is only $2$ TCP segments, which is $536 \times 2$ bytes with IPv4 and $1220 \times 2$ bytes with IPv6, both are smaller than the most coding window sizes we simulated. Our speculation is that the authors of TCP/NC clearly know the potential decoding delay problem with their designed protocol, and limit the coding window size accordingly.

\begin{figure}[thb]
	\begin{center}
		\includegraphics[width=2.5in]{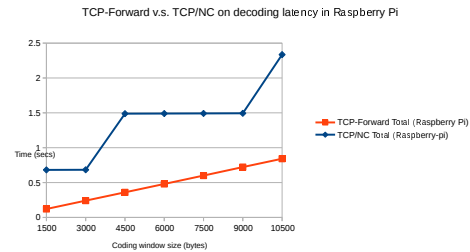}
	\end{center}
	\caption{The decoding delay of both TCP-Forward and TCP/NC on Raspberry Pi. The decoding delay of TCP/NC grows much faster than that of TCP-Forward.}
	\label{fig:tcpforward:tcpnc:pi}
\end{figure}

To further examine the decoding delay in both schemes, we conducted an experiment with 3 hosts (2 PCs and one Raspberry Pi). One PC sends data to the Raspberry Pi which is out of the transmission range. The other PC works as a router between them. Figure \ref{fig:tcpforward:tcpnc:pi} illustrate the decoding delays at the receiver on the Raspberry Pi with the two schemes. As the coding window size increases, TCP-Forward is more resistant to it whereas the decoding delay in TCP/NC grows much faster. Note that after the coding size gets larger than $4500$ bytes, the decoding delay of TCP/NC on a resource limited platform like Raspberry Pi becomes unbearable.

\begin{figure}[hbt]
	\begin{center}
		\includegraphics[width=2.5in]{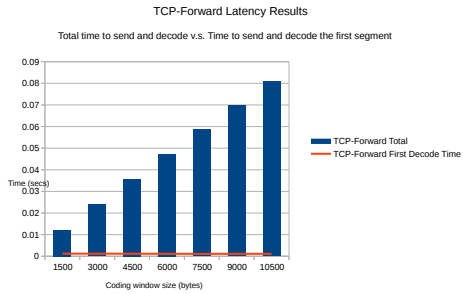}
	\end{center}
	\caption{The latency to encode a coding window and have the receiver successfully decoded all of them, and the latency to encode a coding window and have the receiver successfully decoded the first segment: on a PC.}
	\label{fig:tcpforward:times}
\end{figure}

\begin{figure}[hbt]
	\begin{center}
		\includegraphics[width=2.5in]{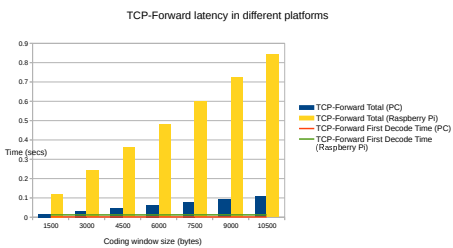}
	\end{center}
	\caption{The latency to encode a coding window and have the receiver successfully decoded all of them, and the latency to encode a coding window and have the receiver successfully decoded the first segment: compare PC to Raspberry Pi.}
	\label{fig:tcpforward:pi}
\end{figure}

One interesting property with Fountain Codes is that, the encoded packets of one generation do not have to be decoded at the same time. There is at least one packet that can get decoded immediately after it is received. Figure \ref{fig:tcpforward:times} shows this property in TCP-Forward. Although the total time to decode every packet in one coding window take longer time, and it grows approximately linearly with the coding window size, the time that it takes to decode the first packet is almost constant and is always short. This is also true in the resource limited system as shown in Figure \ref{fig:tcpforward:pi}, where the total decoding time, and first packet decoding time on both PC and Raspberry Pi are compared. It takes the Raspberry Pi much longer to decode a complete coding window. However, when it comes to the delay from receiving the encoded packets, to at least one packet gets decoded, it is almost as good as the PC.

\section{Discussion}

In this chapter, we give a brief introduction to our newly designed transmission protocol TCP-Forward, with the hope that this protocol can help with solving the TCP's performance problem in the wireless networks. Some experiment and simulation results are presented to demonstrate its performance advantage over not just conventional TCP but also TCP/NC which is a recent TCP variant that introduced Random Linear Network Coding into TCP. Similar to TCP/NC, TCP-Forward tries to keep congestion control algorithms of TCP. We believe congestion control is essentially the most important building block of TCP, and any efforts of replacing TCP without keeping or introducing a new congestion control algorithm is a dead-end.

Compared to TCP/NC, TCP-Forward employs the Fountain Codes which has a clear advantage over Linear Coding. However, incorporating Fountain Codes into a sliding window and a congestion control algorithm is observed to be challenging in our research. Essentially, in a pure FEC setting, there is no need to provide extra redundancy besides a correct redundancy ratio during the encoding process. There does not have to be any data acknowledgement, retransmission, or overlaps from one generation of encoding segments to the next. 
However, completely following the FEC approach is not simple in end-to-end communication. The redundancy ratio may not be perfectly reliable for one reason. The decoding delay from one generation to another can be very long, which leads to premature congestion avoidance mechanism, which breaks the TCP, for another reason.

To handled this problem, TCP/NC introduces a new concept to help the receiver acknowledges a segment even before it gets decoded. Namely the TCP/NC receiver can report it has \textit{seen} a segment even before it is able to decode it. This concept works very well for TCP/NC and helps smoothing the RTT. But, this mechanism cannot be used in TCP-Forward since in Fountain Codes, segments are choose to be included in the encoding process following the Soliton distribution. It is unclear how to modify the \textit{seen} concept to make it work with Soliton distribution. We believe that the half-overlapping sliding window in our current implementation is far from perfect and has left some space of future research for optimization.

That being said, TCP-Forward has already shown its strength. Thanks to the Fountain Codes, TCP-Forward is able to handle a larger coding window without introducing unbearable decoding delays. Different from TCP/NC, we also use an adaptive redundancy ratio during the encoding process. Both these two features help us achieving better throughputs in a lossy network.

\bibliographystyle{IEEEtran}
\bibliography{oracle,rfc}

\end{document}